\documentclass[twocolumn, 3p, number, sort&compress]{elsarticle}

\usepackage[utf8]{inputenc}
\usepackage[T1]{fontenc}

\biboptions{sort&compress,super}

\usepackage{amsmath,amssymb}
\usepackage{graphicx}
\usepackage{balance}
\usepackage{hyperref}
\hypersetup{hidelinks}
\usepackage{csquotes}

\usepackage{mathrsfs} 
\usepackage{bm}
\usepackage{fancyvrb}
\usepackage{color, soul}
\usepackage[usenames,dvipsnames]{xcolor}

\usepackage{multirow}
\usepackage{pdflscape,array,booktabs}

\usepackage{siunitx}
\usepackage{listings}
\newcommand{\tmatrix}{\emph{T}-matrix}

\newcommand{\ninc}{N_\mathrm{inc}}
\newcommand{\nmax}{l_\mathrm{max}}
\newcommand{\lmax}{l_\mathrm{max}}
\newcommand{\sigmacd}{\sigma_\text{\scshape cd}}
\newcommand{\sigmaext}{\sigma_\mathrm{ext}}

\newcommand{\ldoc}{\mathscr{C}}

\title{Orientation-averaged light scattering by nanoparticle clusters:\\ far-field and near-field benchmarks of numerical cubature methods} 

\hypersetup{draft}
\begin{document}

\author{A.~Fazel-Najafabadi}
\ead{atefeh.fazelnajafabadi@vuw.ac.nz}

\author{B.~Auguié\corref{cor1}}
\ead{baptiste.auguie@vuw.ac.nz}

\address{The MacDiarmid Institute for
  Advanced Materials and Nanotechnology,\\ School of Chemical and
  Physical Sciences, Victoria University of Wellington, \\PO Box 600,
  Wellington 6140, New Zealand}
\cortext[cor1]{Corresponding authors}


\begin{abstract}
The optical properties of nanoparticles can be substantially affected by their assembly in compact aggregates. This is a common situation notably for nanoparticles synthesised and self-assembled into rigid clusters in colloidal form, where they may be further characterised or used in spectroscopic applications. The theoretical description of such experiments generally requires averaging the optical response over all possible cluster orientations, as they randomly orient themselves over the course of a measurement. This averaging is often done numerically by simulating the optical response for several directions of incidence, using a spherical cubature method. The simulation time increases with the number of directions and can become prohibitive, yet few studies have examined the trade-off between averaging accuracy and computational cost. We benchmark seven commonly-used spherical cubature methods for both far-field and near-field optical responses for a few paradigmatic cluster geometries: dimers of nanospheres and of nanorods, and a helix. The relative error is rigorously evaluated in comparison to analytical results obtained with the superposition \tmatrix\  method. Accurate orientation averaging is especially important for quantities relating to optical activity, the differential response to left and right circularly polarised light, and our example calculations include in particular far-field circular dichroism and near-field local degree of optical chirality. 
\end{abstract}
\maketitle

\section{Introduction}
%

%


%
\begin{figure}
\includegraphics[width=\columnwidth]{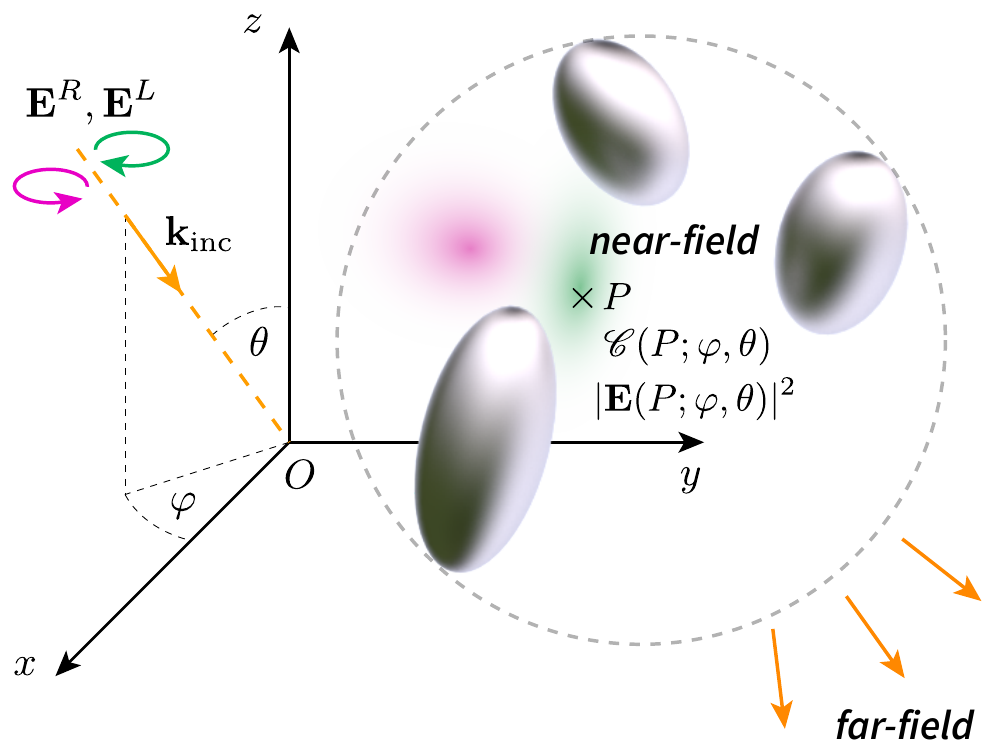}
\caption{Schematic illustration of the light scattering problem under consideration. $N$ nanoparticles, spherical or non-spherical, are placed at arbitrary positions and orientations in a fixed reference frame. This cluster of particles is rigidly held together, but randomly-oriented with respect to incident light (a plane wave with left ($L$) or right ($R$) circular polarisation and incident wavevector $\mathbf{k}_\text{inc}$ oriented by Euler angles $(\varphi,\theta)$ in a reference frame attached to the cluster). We seek the orientation-averaged optical response in the far-field (absorption, scattering, extinction and circular dichroism cross-sections), and in the near-field (local electric field intensity $|\mathbf{E}|^2$ and local degree of optical chirality $\ldoc$) at a specific position $P$, fixed in the cluster's reference frame.} 
\label{fig:fig1}
\end{figure}

Light scattering by nanoparticles underpins many important applications, notably in the realm of optical spectroscopy\cite{Le-Ru:2008wh}, nano-optics\cite{Novotny:2006aa}, and light-harvesting technologies\cite{Zhou:2015wv}. With advances in nanotechnology and synthesis, a wealth of artificial nanostructures have been proposed that combine a number of nanoparticles into rigid aggregates with well-defined positions and orientations in space\cite{Kuzyk:2018uh}, such as, for example, oligomers obtained by top-down lithography\cite{Hentschel:2011vj,Hentschel:2012vj} or helices and other chiral nanostructures obtained by bottom-up assembly\cite{kuzyk2012dna}. The optical properties of such natural or artificial nano-aggregates can reveal a complex interplay between the individual nanoparticles' response, and collective multiple-scattering interactions that depend crucially on the relative positions and orientations of neighbouring particles\cite{Hentschel:2011vj,Lukyanchuk:2010uk,Jain:2006ub,fan2013optical}. Bottom-up synthesis and assembly of nanoparticles in particular enables fine-tuned optical properties, and for such samples the particles or clusters of particles are often fabricated, characterised, and used in colloidal form. Over the typical time-scale of an optical measurement, these compound scatterers in brownian motion assume random orientations, as a whole. 

The superposition \tmatrix\  framework is a powerful method for the theoretical description of light scattering by such aggregates\cite{mishchenko2002scattering, StoutAL02,  Schebarchov:2021wc}; it enables fast and accurate computations of far-field cross-sections as well as near fields\cite{Mackowski:2012ug,Egel:2017we,Schebarchov:2019aa}. A particular strength of the method lies in the prediction of orientation-averaged quantities\cite{mishchenko2002scattering, Avalos-Ovando2021vv, xu2003orientation}: the \tmatrix\  captures the optical response of a scatterer independently of the incident field, and the properties of vector spherical harmonics used to describe incident and scattered fields yield analytical formulas for orientation-averaged optical properties. This powerful formalism enables benchmark calculations for various quantities of interest, which include orientation-averaged extinction, scattering and absorption\cite{mishchenko1990extinction,Khlebtsov92}, circular dichroism\cite{OrientationAveragedMetaMaterials}, but also near-field intensity\cite{StoutTransferMatrixMultipleScatterer2008} and local degree of optical chirality\cite{TangCohen2010,fazel2021orientation}. Although analytical, these orientation-averaged expressions can become quite involved in the case of near-field quantities, requiring the costly evaluation of translation matrices for each evaluation point. As a result, evaluating such analytical formulas is not necessarily faster than performing purely numerical orientation averaging by simulating the optical properties for a discrete number of incidence directions with a numerical cubature\cite{Penttila:2011ws,Okada:2008wf}. This suggests that in many practical applications, where a relative accuracy of $10^{-3}$ is often sufficient in view of other sources of uncertainty, a numerical cubature may be advantageous even when using the \tmatrix\ method. In many other popular electromagnetic simulation methods such as Finite-Difference Time Domain (FDTD)~\cite{taflove2005computational}, Finite-Element Method (FEM)~\cite{volakis1998finite}, Discrete Dipole Approximation (DDA)\cite{Yurkin:2007wv}, the Maxwell equations are solved for a given configuration and a specific incident field, and simulating a new incidence direction incurs almost the full computational cost. For such computational techniques, which do not provide analytical orientation-averaged results and have their own sources of uncertainty, it can be very useful to have accurate benchmark results to compare against, when assessing the accuracy of orientation averaging and deciding on the number of incidence directions to be simulated. We provide below such example computations for several model geometries.

\begin{figure*}[!htpb]
\includegraphics[width=\textwidth]{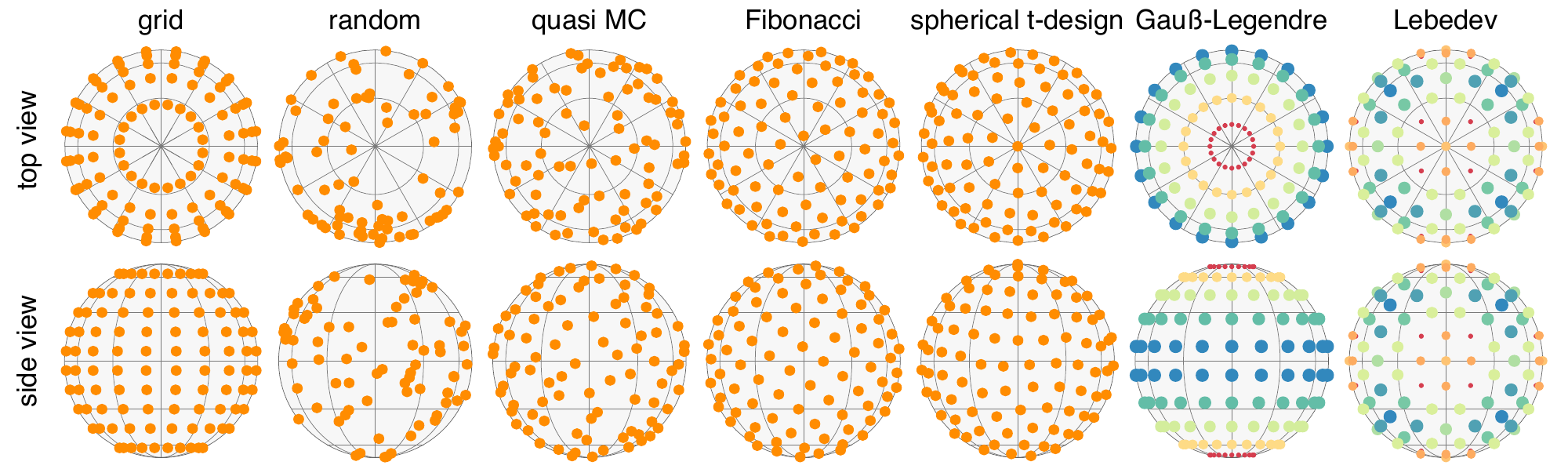}
\caption{Schematic illustration of the 7 spherical cubature rules considered in this work. The nodes $(\varphi_i,\theta_i), i=1\dots \ninc$ are displayed for $\ninc\approx 180$ (exact value differs for some cubature rules, as shown in ESI Fig.~SI1), in 3D spherical polar coordinates (orthographic projections along $z$ (\emph{top view}) and $x$ (\emph{side view})). Gau\ss-Legendre and Lebedev cubatures have non-equal weights, here illustrated with size and colour.} 
\label{fig:fig2}
\end{figure*}

Orientation averaging of optical properties can be defined as an integral over Euler angles $(\varphi,\theta)$, refering to the orientation of the wavevector in spherical coordinates for a plane wave incident on the scatterer (Fig.~\ref{fig:fig1}),
\begin{equation}
\langle f\rangle=\frac{1}{4\pi}\int_0^{\pi} \int_0^{2\pi}  f(\varphi,\theta)  \sin\theta\, \mathrm{d} \varphi \mathrm{d} \theta .
\label{eq:eulerOA}
\end{equation}
where the quantity $f$ corresponds here to an optical response of interest such as far-field cross-sections or near-field intensities at a fixed location $P$ in the scatterer's reference frame. Note that the scatterer is itself a compound object in the case of a particle cluster, and in this work we assume that the component particles are rigidly-held together by a template with negligible influence on optical properties. This averaging over all possible incidence directions is often complemented by a further averaging over polarisation~\cite{mishchenko1989interstellar, mishchenko1990extinction,Khlebtsov92, borghese2007scattering}. In the case of circular polarisation, relevant to the simulation of chiroptical spectroscopy experiments\cite{Berova:2012aa}, we may simply calculate the orientation-averaged result for left ($L$) and right ($R$) polarisations separately, and average both if an unpolarised optical response is sought. Note that this procedure only works for quadratic quantities expressed in terms of electric and magnetic fields, as considered in this work; the calculation of enhancement factors in surface-enhanced Raman scattering requires more care, as some cross-terms are also present~\cite{Khlebtsov:2021vi}.

In practice, the integral in Eq.~\ref{eq:eulerOA} can be approximated by a finite sum using a variety of spherical cubature methods\cite{Penttila:2011ws,Beentjes:2015tz,Hesse:2015ti},
\begin{equation}
\langle f\rangle\approx  \sum_{\varphi_i,\theta_i}^{i=1\dots \ninc} f(\varphi_i,\theta_i) w_i,
\label{eq:cub}
\end{equation}
where the integrand is evaluated at a finite number $\ninc$ of incidence directions $(\varphi_i,\theta_i)$, each value scaled by a corresponding weight $w_i$. Multiple cubature methods have been proposed, differing in their prescription for the nodes $(\varphi_i,\theta_i)$, and associated weights $w_i$ (Fig.~\ref{fig:fig2}).
For the finite sum in Eq.~\ref{eq:cub}, with number of nodes $\ninc$, the accuracy of a given cubature against the analytical result $\langle f\rangle$ can be quantified in the relative error,
\begin{equation}
\varepsilon(\ninc):=\left|\frac{\left(\sum_{\varphi_i,\theta_i}^{i=1\dots \ninc} f(\varphi_i,\theta_i) w_i\right) - \langle f\rangle}{\langle f\rangle}\right|
\label{eq:error}
\end{equation}
(assuming $\langle f\rangle\neq 0$).

We detail below a comparison of seven different numerical cubature methods and their application in orientation averaging of optical properties -- both in the far-field and in the near-field. Their relative performance is first illustrated on toy problems consisting of known integrands, followed by realistic light scattering problems of varying degree of difficulty. We use the analytical results as a benchmark for accuracy, and compare the convergence rate of the different methods with increasing number of incidence angles. 

\section{Spherical cubature}
\begin{figure*}[!htpb]
\includegraphics[width=\textwidth]{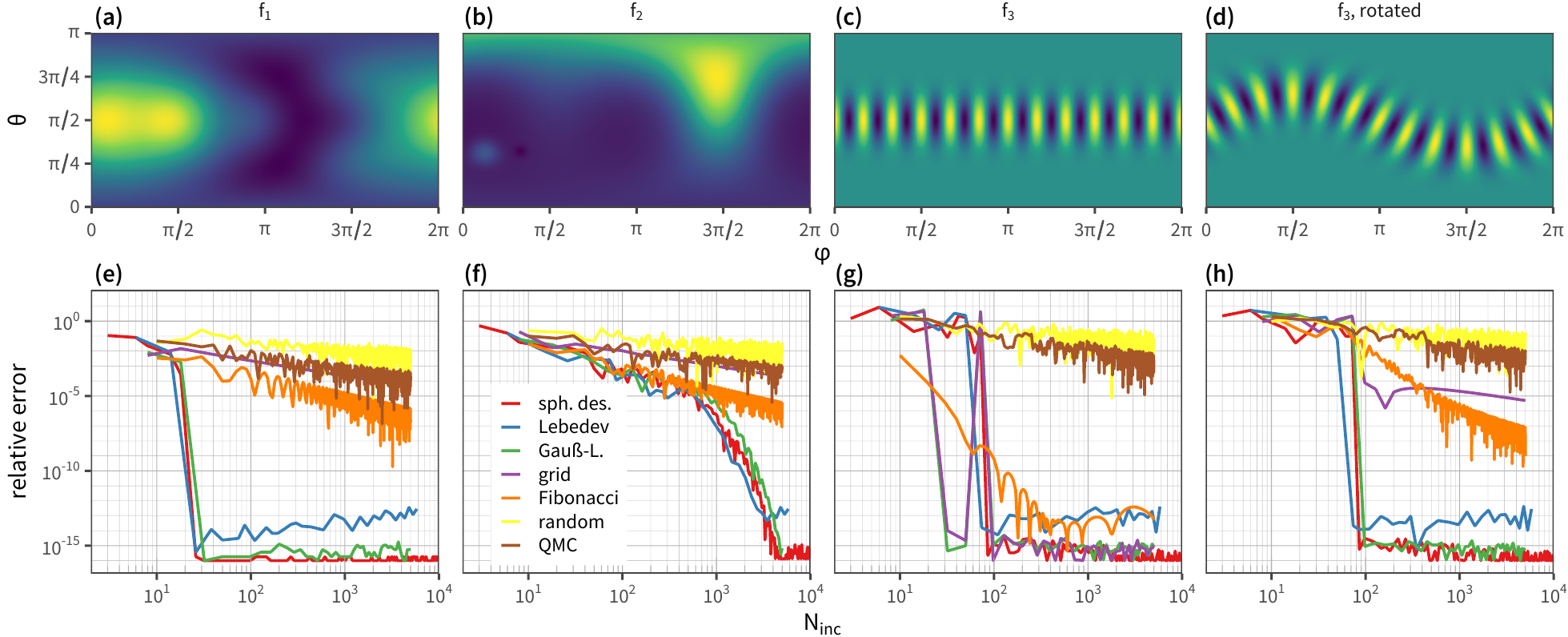}
\caption{Convergence of different spherical cubature rules on 3 analytical integrands. (Top, panels a--d) Colour maps of the integrands, scaled to the same colour range for ease of comparison. Panel (d) presents the same integrand $f_3$ as panel (c), rotated by an arbitrary angle $\theta=27.1$ degrees. (Bottom, panels e--h) Convergence plots in log-log scale for 7 spherical cubature methods. The relative error is defined in Eq.~\ref{eq:error} in reference to the known analytical result for each integral: $\langle f_1\rangle = 216\pi/35$, $\langle f_2\rangle = 6.6961822200736179523\dots$, $\langle f_3\rangle = 1$.} 
\label{fig:fig3}
\end{figure*}

We implemented seven well-known methods of spherical cubature, namely (i) \emph{`grid'} -- a cartesian product of mid-point rules in $(\varphi, \cos\theta)$, ; (ii) \emph{`random'}, also known as Monte Carlo -- a random sampling of points in $(\varphi, \cos\theta)$; (iii) \emph{`QMC'} -- a quasi Monte Carlo sampling over $(\varphi, \cos\theta)$ based on a Halton sequence\cite{Penttila:2011ws}, (iv) \emph{`Fibonacci'} -- a relatively uniform spiral covering of the sphere\cite{Penttila:2011ws}; (v) \emph{`spherical t-design'}\cite{Womersley} ; (vi) \emph{`Gau\ss-Legendre'} -- a cartesian product of Gau\ss-Legendre 1D quadrature for $\cos\theta$, and a mid-point rule for the azimuth $\varphi$ ; and (vii) \emph{`Lebedev'}\cite{lebedev1976quadratures}. More details are given in ESI Section S1 regarding the implementation of these methods, which we also make available in the \texttt{R} package "\texttt{cubs}"\cite{cubs}. Figure \ref{fig:fig2} depicts the angular distribution of $\ninc\approx 180$ nodes for each cubature, noting that the exact number of nodes varies with each method, some having coarser steps between possible values (this granularity is presented in ESI Fig.~SI1). 

In order to validate the implementation of each cubature method, we first tested their accuracy against analytical cases, following Ref.~\cite{Fornberg:2014tn,Beentjes:2015tz} (Fig.~\ref{fig:fig3}),

\begin{flalign}
f_{1}(x, y, z)=& 1+x+y^{2}+x^{2} y + x^{4}+\notag  \\
 &  + y^{5}+x^{2} y^{2} z^{2},
\end{flalign}
\begin{flalign}
f_{2}(x, y, z)=& \tfrac{3}{4} e^{\left[-(9 x-2)^{2} / 4-(9 y-2)^{2} / 4-(9 z-2)^{2} / 4\right]}\notag \\
&+\tfrac{3}{4} e^{\left[-(9 x+1)^{2} / 49-(9 y+1) / 10-(9 z+1) / 10\right]}\notag \\
&+\tfrac{1}{2} e^{\left[-(9 x-7)^{2} / 4-(9 y-3)^{2} / 4-(9 z-5)^{2} / 4\right]}\notag \\
&-\tfrac{1}{5} e^{\left[-(9 x-4)^{2}-(9 y-7)^{2}-(9 z-5)^{2}\right]},
\end{flalign} 
with the usual spherical coordinates,
\begin{flalign}
x = &\cos(\varphi)\sin(\theta)\\
  y = & \sin(\varphi)\sin(\theta)\\
  z = & \cos(\theta),
\end{flalign}
and a spherical harmonic
\begin{flalign}
f_{3}(\varphi, \theta)=& \tfrac{1}{4\pi} + \cos(12\varphi)\sin^{12}(\theta),
\end{flalign}
where we added a constant to ensure the denominator of Eq.~\ref{eq:error} does not vanish.

The performance of the seven cubature methods differs markedly on these examples, and we distinguish two families: `naive' or `generic' methods comprising \emph{`grid'}, \emph{`random'}, \emph{`QMC'}, \emph{`Fibonacci'}, which show poor convergence and reach a limited accuracy for a few hundreds points; in contrast, the three methods \emph{`Gau\ss-Legendre'}, \emph{`Lebedev'}, and \emph{`spherical t-design'} show much faster convergence, and reach excellent accuracy in double precision arithmetic for these three examples. The relative error for these methods presents an abrupt drop before plateauing at near-maximum precision after a certain number $\ninc$, the value of which depends on the integrand. $f_1$ only requires $\ninc\sim 20$ to reach this plateau, while $f_2$ requires $\ninc\sim 4000$, which we can attribute to the increased angular complexity of the integrand. This aspect can be quantified more formally by considering a spherical harmonic (SH) decomposition of the integrand, and comparing the relative magnitude of each spherical harmonic coefficient\cite{Fazel-Najafabadi:2021ud}, as we discuss further below. The example of $f_3$ corresponds to a pure spherical harmonic, namely $Y_{12}^0$, with a convergence falling in-between the two previous cases. We note that integrands with high symmetry, such as $f_3$, can prove misleading in such comparisons: a regular `grid' method, or the Fibonacci sequence, for example, here appear to approximate accurately the integral at low $\ninc$ values, but only because the sampling of points on a regular grid coincides with a symmetry of the integrand. This artificial coincidence is removed in the right-most panel, where the same integrand $f_3$ is now rotated by a non-trivial angle $21.7^\circ$ about the z-axis. The  coincidental sampling of $f_3$ at symmetrical points, leading to cancellations in the cubature, no longer occurs, and both `grid' and `Fibonacci' methods return to slow convergence rates. In contrast, \emph{`Gau\ss-Legendre'}, \emph{`Lebedev'}, and \emph{`spherical t-design'} are unaffected by this rotation. Rotating a spherical harmonic $Y^m_l$ mixes different angular momentum ($m$) values, but only within a given degree, and $f_3$ is therefore transformed into a weighted sum of spherical harmonics $Y_{12}^m, m=-12\dots 12$. These three cubature methods are known to integrate exactly all spherical harmonics up to a specific degree, regardless of $m$. For an integrand composed of degree $\leq 12$ spherical harmonic(s), the plateau is reached at $N_\text{inc}=74$ (Lebedev), $N_\text{inc}=86$ (spherical t-design), and $N_\text{inc}=98$ for Gau\ss-Legendre. Although Gau\ss-Legendre is somewhat less efficient than the other two, and has fewer available numbers of points, it may be the preferred method in configurations with a known symmetry -- for example where only one octant of the full solid angle needs to be considered, or if the integrand is invariant in $\varphi$ -- in which case the choice of nodes in $\theta$ and $\varphi$ may be adapted accordingly, with more freedom than for the Lebedev and spherical t-design methods.

These examples were chosen for demonstration purposes, but our interest is to approximate the integral of light-scattering quantities, which generally admit no analytical expression. The integrand will be computed numerically with the superposition \tmatrix\ method\cite{Schebarchov:2021wc}, and particular configurations can yield very differently-behaved integrands. Typically, we expect that a compact, small-sized cluster (compared to the wavelength) will show a relatively featureless angular pattern, and spherical cubatures of small order will suffice for an accurate orientation averaging process. Indeed, averaging the response along $x$, $y$, $z$ axes is a commonly-used procedure for orientation-averaging in this regime\cite{fan2013optical}. Larger particles and/or clusters of particles may however show very directional responses, such as the photonic jet effect in large Mie scatterers\cite{Lukyanchuk:2017uh}, or waveguide modes along chains of particles\cite{Fazel-Najafabadi:2021ud} or dielectric particles with a high aspect ratio\cite{Kerker:1980va}. 

An important question we set out to investigate is the variation in orientation averaging difficulty, namely the number of incidence angles required to reach a given accuracy, with the light scattering quantity of interest. Previous studies have been devoted to spherical cubature for far-field cross-sections\cite{Penttila:2011ws}, but it is not clear if the conclusions are equally-applicable to near-field quantities; in fact, the general expectation is that the near-field will be more challenging, as it typically involves higher-order multipoles in the \tmatrix\ framework (or Mie theory, for single spheres). We also hypothesise that chiroptical properties, namely the differential response to left and right circularly polarised light, in both far-field and near-field settings, may prove more challenging for orientation averaging than unpolarised responses. The following examples were chosen to explore these questions on several cluster geometries of interest.


%
\section{Dimer}
\begin{figure}[!htpb]
\includegraphics[width=\columnwidth]{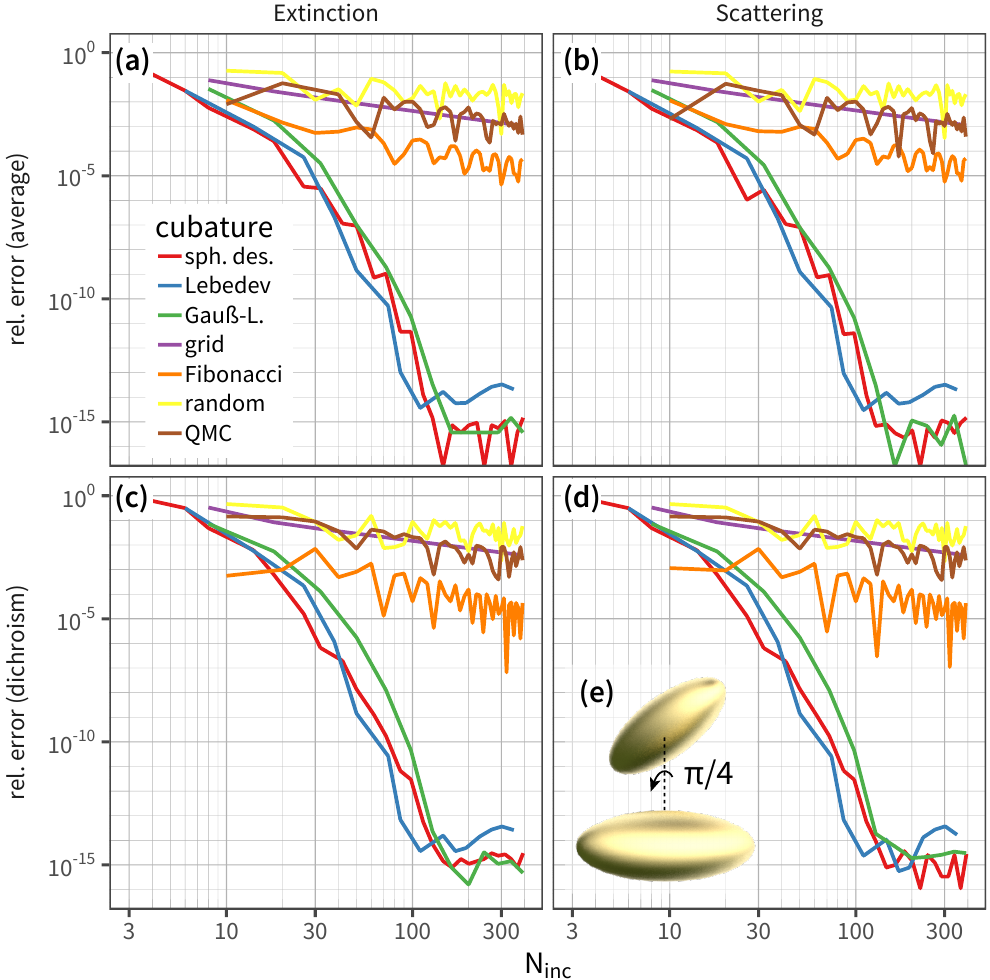}
\caption{Orientation averaging of light scattering by a chiral dimer of gold nanorods immersed in water ($n=1.33$). The two nanorods are modelled as prolate spheroids with semi-axes $a=b=30$\, nm, $c= 80$\, nm; with centre-to-centre separation $d=200$\,nm, and dihedral angle $\varphi=\pi/4$ about the dimer axis. (a--b) Convergence of the far-field extinction (a) and scattering (b) cross-section at wavelength $\lambda=800$\,nm, for different cubature rules. (c--d) Corresponding convergence of circular dichroism. (e) Schematic view of the dimer structure.} 
\label{fig:fig4}
\end{figure}
\begin{figure*}[!htpb]
\includegraphics[width=\textwidth]{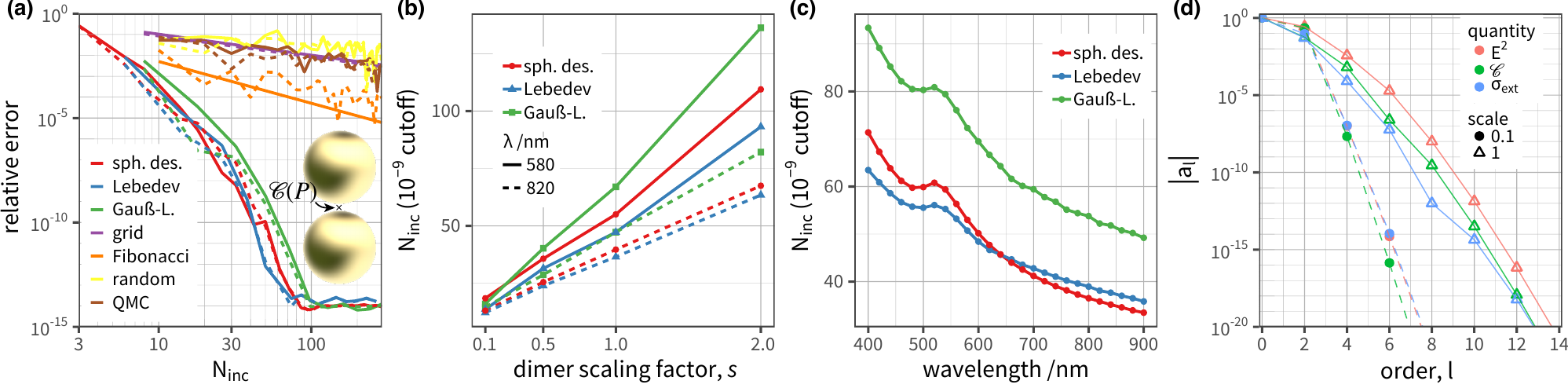}
\caption{Orientation averaging of light scattering by a dimer of gold spheres, as depicted in panel (a). The nominal dimensions are $r=50$\,nm for the sphere radius, and a gap $g=10$\,nm. We later vary both dimensions simultaneously with a scaling factor $s$. The dashed lines correspond to the results for a dimer tilted by Euler angles $\varphi=12^\circ,\theta=21^\circ$ from the $z$ axis. (a) Convergence of 7 cubature rules for the local degree of chirality $\ldoc$ at the mid-gap point $P$ between the spheres. The scaling factor is $s=1$ in this plot. (b) Variation of the number of cubature angles required to reach an accuracy of $10^{-9}$ as a function of the dimer's scaling factor, $s$. (c) Variation of the number of cubature angles required to reach an accuracy of $10^{-9}$ as a function of the wavelength, for the nominal dimer ($s=1$). (d) Spherical Harmonic Decomposition of the angular pattern for far-field extinction (blue symbols), near-field intensity $|\mathbf{E}|^2$ (pink symbols), and local degree of chirality $\ldoc$ (green symbols), for 2 scale factors ($s=0.1$, solid circles, and $s=1$, open triangles). The variable $l$ refers to the order of spherical harmonic.} 
\label{fig:fig5}
\end{figure*}
Our first illustration considers light scattering by a `fingers crossed' dimer of gold nanorods\cite{auguie2011fingers}, which has attracted considerable interest in recent years as a prototypal structure for chiral plasmonics\cite{yin2013interpreting, lan2013bifacial, ma2013chiral, yin2015active,zhou2015plasmonic,  najafabadi2017analytical,  zhao2017chirality,jiang2017stimulus}. We recently revisited this geometry with a focus on the angular dependence of the chiroptical far-field properties, namely the circular dichroism in extinction, scattering and absorption\cite{Fazel-Najafabadi:2021ud}. Despite its simplicity, involving only two particles, the structure reveals an interesting response arising from the hybridisation of localised plasmon resonances, and a balance between absorption and scattering contributions as the size of the structure is varied\cite{auguie2011fingers}. The angular response is non-trivial: for example, light incident normal to the dimer axis exhibits no circular dichroism. For very small dimers this angular response is relatively smooth nonetheless, and as a result the orientation-averaged response may be reasonably approximated by the average over 3 orthogonal incidence directions. As the scale of the dimer increases, however, this is no longer an accurate approximation\cite{Fazel-Najafabadi:2021ud}. We confirm here this observation with a more complete characterisation of numerical angular averaging, using the previously-defined seven cubature methods (Fig.~\ref{fig:fig4}). The size of the dimer is chosen such that the structure extends to a sizeable fraction of the wavelength at resonance, i.e. beyond the Rayleigh regime. In agreement with our earlier observation, the same three cubature methods stand out (Lebedev, spherical t-design, Gau\ss-Legendre), while the others show a slow convergence with a modest relative accuracy of $\sim 10^{-4}$ requiring hundreds of incidence angles. For the three best-performers, this number is reduced to a few dozen angles, much more practical with time-consuming numerical methods. Figure~\ref{fig:fig4} distinguishes between scattering and extinction (scattering + absorption) cross-sections, but we find a very similar convergence behaviour. Similarly, we considered separately unpolarised cross-sections and their corresponding circular dichroism (differential response between left and right polarisations), but find that the convergence properties with respect to orientation averaging are very similar in both cases.

%

Our next example aims to further clarify the difficulty of orientation averaging, i.e. the number of incidence angles required for an accurate cubature, in relation to the geometry of the cluster. In Fig.~\ref{fig:fig5} we consider a dimer of gold spheres -- the simplest multi-particle cluster --, and vary the scale of the structure by a factor $s$ (sphere radius and inter-particle distance are scaled identically). For a given value of $s$, we obtain convergence results as previously discussed (Fig.~\ref{fig:fig5}(a) for $s=1$, showing the convergence of $\ldoc$ at the mid-point $P$ between the two spheres, but other quantities reveal a similar trend). We also verify that tilting the dimer away from the $z$ axis (dotted lines in Fig.~\ref{fig:fig5}(a)) does not affect our conclusions; some cubature methods such as Fibonacci show a more unpredictable convergence with the reduced symmetry, but the trends are very similar.

The absolute precision of orientation-averaged results depends not only on the number of incidence angles, but also on the other parameters of the simulation. The uncertainties in physical parameters, such as particle size and shape distribution, material properties, etc. often limit the agreement between theory and experiment to much lower accuracy in many contexts\cite{Grand:2019ty}. The computational method itself is also subject to other sources of error, such as the fineness of a mesh when the scatterers are discretised in FEM, FDTD or DDA; in the \tmatrix\ method employed here, the key convergence parameter is the maximum multipolar order $\lmax$ which is used to truncate field expansions\cite{MishchenkoTL02}. In our implementation, note that $\nmax$ refers to scatterer-centred expansions\cite{StoutTransferMatrixMultipleScatterer2008,Schebarchov:2021wc}. Remarkably, we find no interaction between $\ninc$ and $\lmax$ when studying the convergence of orientation-averaged quantities. For example, varying $\lmax$ from 3 to 20 does not affect the convergence of spherical cubature (ESI Fig.~S2). The net accuracy of the results is of course lower at lower $\lmax$ -- it may be just $10^{-3}$ (ESI Fig.~S3) while the cubature still reaches a plateau below $10^{-13}$ in the agreement between numerical cubature and analytical orientation averaging. This suggests that the analytical formulas for orientation averaging are also valid at finite $\lmax$, not requiring the limit $\lmax\to\infty$ as some derivations would seem to imply. Conveniently, this absence of interaction between $\ninc$ and $\lmax$ also means that we can reliably test cubature convergence at relatively small $\lmax$ values. 

In practice, many light-scattering calculations do not require the accuracy achieved by the Levedev, Gau\ss-Legendre, or spherical t-design cubatures at the level where they plateau near double-precision accuracy; instead a relative accuracy of $10^{-3}-10^{-5}$ may often be sufficient for most applications, and these cubature methods reach such precision with far fewer evaluations than the other methods. Taking an arbitrary cutoff value, $\varepsilon = 10^{-9}$ for instance, we may estimate the number of cubature points required by each method by simple interpolation, and trace its dependence on simulation parameters such as the scale factor $s$ (Figure~\ref{fig:fig5}(c)), or the wavelength (Figure~\ref{fig:fig5}(d)). For clarity, we only consider from now on the three best-performing cubature methods. Additionally, one should note that the estimated cutoff value may not be exactly obtainable for a given cubature method; we are however interested in the general trends of this cutoff number. 

The results clearly demonstrate that the size parameter of the cluster is key to determining the accuracy of orientation averaging. The cutoff number increases linearly with the scale of the compound scatterer, at a given wavelength, with the slope varying slightly from one method to another. An inverse dependence on wavelength is also apparent, although the variation of the material's dielectric function with wavelength introduces further dependencies, as observed in the peak appearing around 520\,nm, associated with a plasmon resonance. The inter-particle coupling strength generally depends on material properties as well as the geometry, and this in turn affects the relative strength of different multipolar orders to describe the scattering of the whole cluster. 

We performed a spherical harmonic (SH) decomposition of the angular pattern of $f(\varphi,\theta),\quad f\in \left\{\sigma_\mathrm{ext},|\mathbf{E}|^2, \ldoc\right\}$ for clusters of two different scales: $s=0.1$ and $s=1$ (Fig.~\ref{fig:fig5}(d)). This decomposition of the integrand into a series of spherical harmonics performs a similar task as Fourier transforms for one-dimensional signals, but on the surface of the unit sphere. The SH decomposition expands the function $f(\varphi,\theta)$ as
\begin{equation}
f(\varphi,\theta) \approx \sum_{l=0}^{l_\mathrm{max}} \sum_{m=-l}^l a_{lm} Y_l^{m}(\varphi,\theta),
\label{eq:sh}
\end{equation}
with $Y_l^{m}$ the standard spherical harmonics of degree $l$ and angular momentum $m$. Higher-order spherical harmonics are associated with more localised angular features, which in turn require more angles of incidence for numerical cubature. We used the Matlab \texttt{Chebfun} library\cite{T.-A.:2014wj,chebfun:2019aa} for this decomposition, which implements internally an efficient algorithm based on Fourier transforms. Note that we collapse the coefficients into $|a_l|:=\sum_{m=-l}^l |a_{lm}|^2$, to summarise the relative weight of a given order $l$, and normalise the coefficients so that $|a_0|=1$ in each case. The contrast between the two cluster scales in Fig.~\ref{fig:fig5}(d) is very clear, and consistent across several optical quantities of interest (far field extinction cross-section, $\sigmaext$, local field intensity $E^2$, and local degree of optical chirality $\ldoc$). The magnitude of the SH coefficients rapidly decays with order $l$, but this decay is much faster for the smaller cluster. We also note that $E^2$ and  $\ldoc$ show a slower decay than the far-field cross-section, suggesting that near-field quantities are indeed more challenging to integrate numerically. This is the focus of our next example, with a more complex structure.

\section{Helix}
\begin{figure}[!htpb]
\includegraphics[width=\columnwidth]{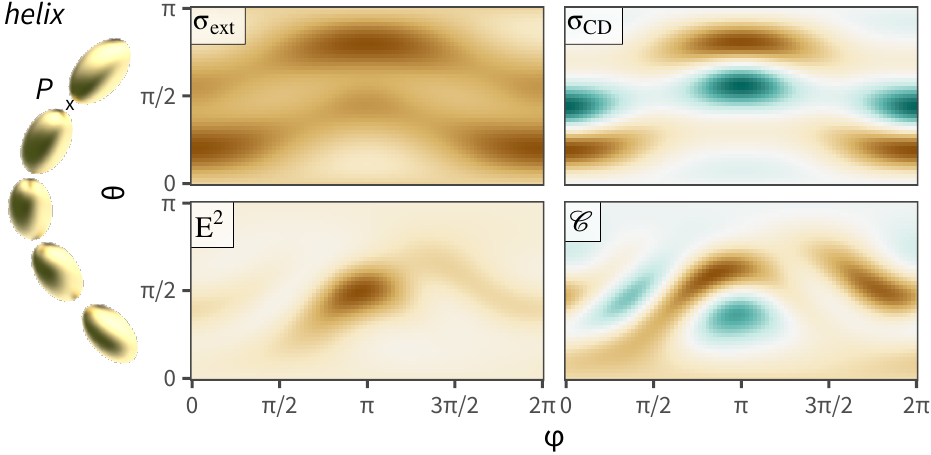}
\caption{Orientation dependence of light scattering by a helix of gold nanorods (left image). The cluster consists of five prolate gold spheroids, with helix axis $z$, radius $\SI{100}\nm$, pitch $\SI{700}\nm$, inter-particle angular step $\delta = \pi/4$. The spheroids are oriented along the helix, and have semi-axes $a=b=\SI{30}\nm$ and $c = \SI{50}\nm$. The colour maps present the angular pattern of far-field cross-sections $\sigmaext$ and $\sigmacd$, and near-field values $|\mathbf{E}|^2$ and $\ldoc$ at location $P$, half-way between the last two spheroids. The data for each panel are normalised to their maximum value (brown, with white set to 0 and green corresponding to negative values), as we are interested in comparing the angular patterns rather than specific numeric values (which also have different units). These simulations used $\nmax=15$.} 
\label{fig:fig6}
\end{figure}

The previous examples considered compact clusters, with just two particles. With more particles, the optical response can become even richer, as the resonances of individual particles hybridise and form collective modes. These larger clusters typically require higher multipolar orders to describe the overall response of the cluster to incident light about a single origin, resulting in a more intricate angular pattern and a more challenging orientation averaging via numerical cubature. We illustrate such a situation by considering a helical strand of gold nanorods previously studied\cite{fazel2021orientation}, similar in size to some recently-proposed self-assembled structures\cite{severoni2020plasmon}. 

\begin{figure}[!htpb]
\includegraphics[width=\columnwidth]{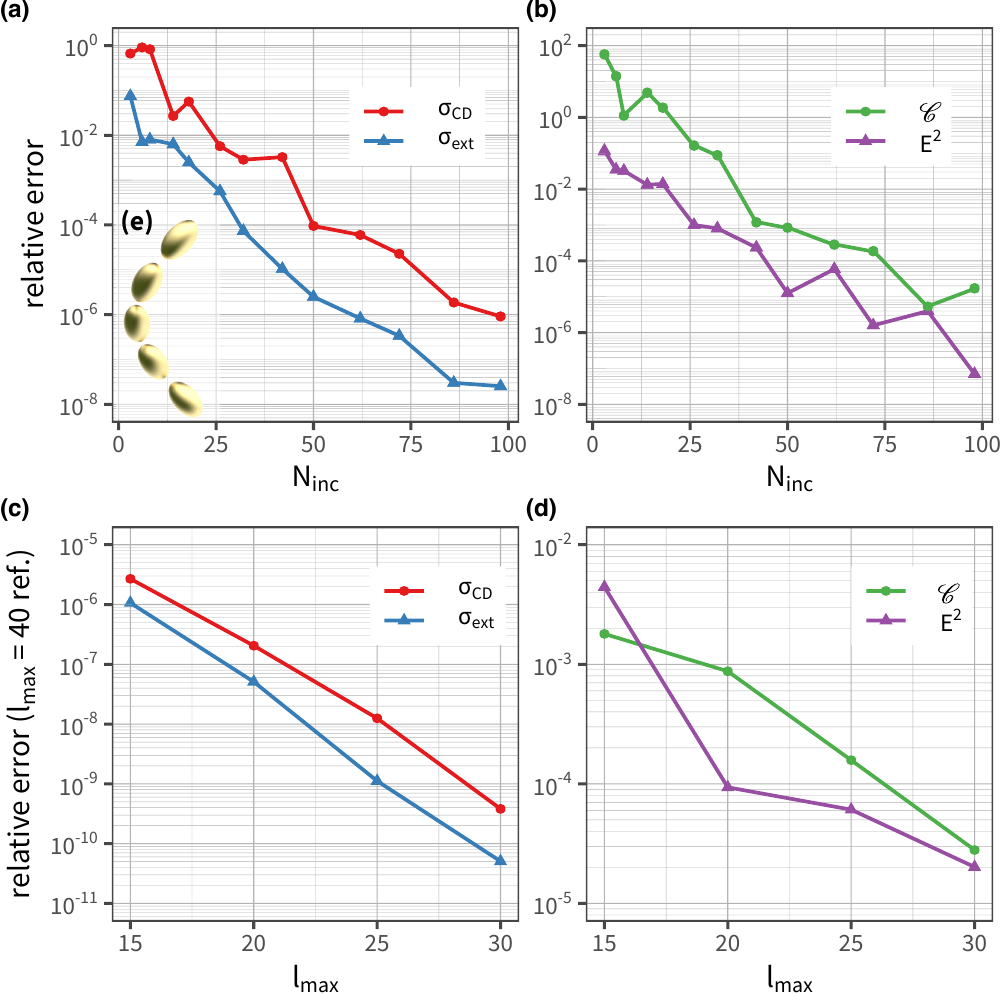}
\caption{Orientation averaging of light scattering by the helical structure of Fig.~\ref{fig:fig6}. (a) Convergence of the relative error for far-field cross-sections (extinction $\sigmaext$, blue, and corresponding circular dichroism $\sigmacd$, red). (b) Convergence of the relative error in near-field quantities at location $P$ (local field intensity $|\mathbf{E}|^2$, purple, and local degree of chirality $\ldoc$, green). The results are shown for the average over both $L$ and $R$ incident polarisations. (c) and (d) Convergence of the relative error in the same quantities, with respect to $\nmax$. The quantities are here calculated only via analytical orientation averaging.} 
\label{fig:fig7}
\end{figure}
The structure consists of 5 gold spheroids immersed in water (Fig.~\ref{fig:fig6}) and we compute both far-field and near-field quantities, to contrast their orientation dependence. Specifically, we calculate four quantities of interest: the far-field extinction cross-section $\sigma_\text{ext}$, the corresponding circular dichroism $\sigma_\text{\scshape cd}$, the near-field local intensity $|\mathbf{E}|^2$ at a position $P$ mid-way between the last two particles, and the local degree of optical chirality $\ldoc$ at the same position. Full spectra are shown in ESI Fig.~S4, and in the following we focus on the wavelength $\lambda=650$\,nm. The angular dependence of each quantity over the full solid angle is presented in Figure \ref{fig:fig6}, where the colour maps are scaled to a maximum value of 1 for ease of comparison. The patterns are non-trivial, as the structure is relatively extended, and sharp angular features are present. They differ between far-field and near-field, with the local field intensity presenting a single maximum around ($\varphi=\pi,\theta=\pi/2$) for this particular position $P$, while the far-field patterns have several maxima of equal strength. The chiroptical properties show more complex patterns than their unpolarised equivalent, as they can take both positive (brown) and negative values (green).

We now turn to the convergence behaviour of orientation averaging for these four quantities of interest (Fig.~\ref{fig:fig7}). For clarity we only present results obtained with spherical t-design cubatures, which showed excellent accuracy in the previous examples, and provide finer steps between available cubatures than Lebedev. The convergence of far-field properties reveals a marked difference between $\sigmaext$ and $\sigmacd$, with a relative error typically a factor of 10 worse for circular dichroism. This increased difficulty in computing chiroptical properties is not unexpected, and follows from the more complex pattern observed in Fig.~\ref{fig:fig6}. 
The near-field convergence is displayed with a different scale in Fig.~\ref{fig:fig7} (b), as the range of convergence is noticeably poorer than for far-field properties. This is also unsurprising, as the near-field quantities present sharper angular features in Fig.~\ref{fig:fig6}. The local degree of chirality is also showing a worse convergence rate compared to the local field intensity, confirming that chiroptical properties are more challenging to integrate accurately than unpolarised ones.

As noted previously for the dimer structure, the convergence of the results with respect to the number of incidence angles is largely independent of the multipole truncation parameter $\lmax$ used in the simulations. This value however dictates the net accuracy of the calculations, and we show in Fig.~\ref{fig:fig7}(c,d) the convergence with respect to $\lmax$. Here the exact result, corresponding formally to $\nmax=\infty$, is not known, and we use $\nmax=40$ as reference instead. Similar conclusions can be drawn, with near-field quantities converging much slower than far-field cross-sections, and chiroptical properties being more challenging to calculate than unpolarised ones in both near-field and far-field settings. These conclusions are expected to hold beyond this particular set of examples, except for particular cases such as near-field points of high-symmetry.



%
\section{Discussion and conclusions}

Among the important decisions to consider when simulating optical properties of nanoparticles or nanoparticle clusters is the choice of the incident light, often taken as a plane wave, notably its state of polarisation and direction of propagation. The optical properties can differ dramatically along different directions, and in the context of chiroptical properties, may change the response entirely with the onset of so-called extrinsic chirality\cite{Plum:2009ua}. Many popular numerical methods such as FDTD, FEM, or DDA, require a complete new calculation for each direction of incidence, with a substantial cost in simulation time. Although modern solvers of Maxwell's equations are getting increasingly performant, and the computational hardware allows large-scale problems to be treated on standard desktop computers, the computational cost is by no means negligible and remains a hindrance for large-scale optimisation problems. Indeed, a cluster of just a few particles already exposes many degrees of freedom, particularly for nonspherical particles, or if their geometry itself can be modified\cite{Yao:2021vp}. Additionally, the quantities of interest often involve further averaging, such as enhancement factors averaged over a specific volume or surface area where analytes may be present\cite{Khlebtsov:2021vi}, or the integration over a given spectral range\cite{Johnson:2019tk}. Real-world particles are typically polydisperse, requiring further averaging over the size and shape distribution to compare quantitatively with experiments. With each new variable of integration the problem becomes less tractable with off-the-shelf computers, thereby limiting the number of studies and the full exploration of such rich parameter spaces. In this context, orientation averaging is an easy question to address: for randomly-oriented scatterers, one typically wishes to obtain a reliable estimate of the orientation-averaged quantity of interest, and often a relative accuracy of $10^{-3}$ will be sufficient, taking into account the often larger sources of uncertainty in nanoscale dimensions, morphology, or even in the material composition and dielectric function\cite{Djorovic:2020vv}. With this objective in mind, one hopes to find \emph{rules of thumb} to decide a priori on a reasonable number of incidence directions to simulate, as well as a means to evaluate the magnitude of the error in this averaging process. The situation is however complicated by the fact that different quantities of interest have different convergence behaviour, and few studies have been devoted to the performance of different cubature rules in light scattering simulations.


Our results offer some progress toward these goals, with three notable contributions: first, we benchmark seven different spherical cubature methods, considering both far-field and near-field quantities of interest; second, we also contrast the standard unpolarised quantities such as cross-sections and field enhancement, with chiroptical properties including circular dichroism and local degree of optical chirality, shown to display a more challenging angular dependence; third, we connect the convergence of cubature rules to the spherical harmonic decomposition of the integrand.

While no universal rule stands out to offer an a priori estimate of the number of incidence directions needed for a given nanostructure, we have identified some helpful indicators. First and foremost, the particle cluster's size parameter, i.e. the radius of its circumscribed sphere times the wavenumber in the incident medium, dictates the variability of the angular pattern, and consequently, the degree of spherical cubature required for an accurate integration. This necessitates using one of the best-performing spherical cubature methods, namely Lebedev, spherical t-design, or Gau\ss-Legendre rules. These methods integrate exactly spherical harmonics up to a given order, and we confirmed their superior properties over `generic' methods such as regular grids, quasi Monte Carlo, or Fibonacci, which nonetheless remain popular choices. Inter-particle coupling depends on many factors, from the exact geometry of the particle cluster, but also the resonances supported by each particle. As a result, the wavelength dependence of the convergence properties depend to an extent on the material properties, not just on the wavenumber. 

We also observed a general trend when comparing unpolarised quantities of interest with chiroptical properties, in both the far-field and in the near-field. Chiroptical properties are found more sensitive to the direction of incidence, with more complex angular patterns and a need for higher-order cubatures to reach a given relative accuracy.
Similarly, we contrasted near-field and far-field quantities of interest, where, as expected from empirical knowledge, we find that the convergence of near-field quantities is more computationally-demanding. Near-field properties typically require taking into account multipolar components or higher order, compared to far-field properties. Remarkably, however, we find that the convergence rate of far-field or near-field quantities is largely independent of the multipolar truncation cutoff used in the \emph{T}-matrix simulation: although the actual accuracy of the results is affected, the relative accuracy of numerical cubature against analytical formulas shows a consistent convergence with respect to the number of cubature angles. This unexpected observation suggests that the analytical formulas for orientation-averaged properties provide robust estimates even at relatively low multipole order, even though their original derivation considers the results only strictly valid in the limit $\nmax\to\infty$. A practical benefit of this observation is that one may carry out a cubature convergence study at relatively low $\nmax$ to decide on the number of incidence directions (with the analytical benchmark using the same $\nmax$ value),  and this cubature can then be applied to the final -- more time-consuming -- calculation at higher $\nmax$.

We note that the spherical cubature methods employed in this work are in some way suboptimal, compared to adaptive cubature which can subdivide the unit sphere recursively until the relative accuracy in each subsection is below a minimum threshold. Such strategies are potentially very advantageous, as they \emph{adapt} to each particular integrand, refining the sampling of incidence angles in the regions of stronger variation. In contrast, the cubature methods used in this work have predetermined nodes and weights, regardless of the integrand's angular profile. The advantage is expected to be highest for sharply-defined angular properties, as exhibited by large elongated nanostructures supporting collective modes with well-defined momentum\cite{Fazel-Najafabadi:2021ud}. We tested briefly some adaptive cubature methods\cite{Johnson:wh,Nolan:2021uh}, and did not observe a better performance on the above examples compared to Levedev, spherical t-design, or Gau\ss-Legendre cubatures. This is however a promising extension of this work. Similarly, it would be interesting to consider nested cubature rules, similar to Gau\ss-Kronrod for 1D quadrature, which would provide an error estimate together with the integral\cite{Press:2007te}.



%


%
\section*{Acknowledgements}
The authors would like to thank Dmitri Schebarchov and Eric Le Ru for helpful discussions, the Royal Society Te Ap\=arangi for support through a Rutherford Discovery Fellowship (B.A.), and the MacDiarmid Institute for additional funding (A.F.N.).

\bibliographystyle{elsarticle-num}
\bibliography{references}

%

%
\end{document}